\documentclass[pra,showpacs,twocolumn]{revtex4}
\usepackage{dcolumn}
\usepackage{graphicx}\usepackage{amsmath}
\newcommand{\opr}[1]{\mathaccent 94 #1}

\newcommand{\avr}[1]{\left\langle #1 \right\rangle}
\begin{document}
\sloppy
\title{Simulation of tunneling in the quantum tomography approach}
\author{Yu. E. Lozovik}
\email{lozovik@isan.troitsk.ru}
\author{V. A. Sharapov}\email{vladsh@e-mail.ru}
\author{A. S. Arkhipov}\email{antoncom@id.ru}
\affiliation{Institute of Spectroscopy RAS, Moscow region, Troitsk, Russia, 142190}
\date{\today}
\begin{abstract}
The new method for the simulation of nonstationary quantum processes is proposed. The method is based on
the tomography representation of quantum mechanics, {\it i.e.}, the state of the system is described by the
{\it nonnegative} function (quantum tomogram). In the framework of the method one uses the
ensemble of trajectories in the tomographic space to represent evolution of the system (therefore
direct calculation of the quantum tomogram is avoided). To illustrate the method we consider the
problem of nonstationary tunneling of a wave packet. Different characteristics of tunneling, such
as tunneling time, evolution of spatial and momentum distributions, tunneling probability are
calculated in the quantum tomography approach. Tunneling of a wave packet of composite particle,
exciton, is also considered; exciton ionization due to the scattering on the barrier is analyzed.
\end{abstract}
\pacs{03.65.Wj, 02.70.-Ns, 03.65.Xp}
\maketitle

\section{Introduction}\label{Introduction}
Nowadays simulation of quantum systems is developed to a high extent (see, {\it e.g.}, reviews
\cite{Lee,Cep}). However, common simulation methods, for example, Path Integral Monte Carlo
or Wigner dynamics, use non-positively defined functions (wave function, the Wigner function,
{\it etc.}) to describe a quantum state. This leads to the difficulties with convergence of
corresponding integrals, especially harmful for the simulation of Fermi systems (sign problem).
There is a hope that using a real nonnegative function, describing the quantum state,
one can avoid these difficulties.

The real nonnegative function in the phase space, completely describing the quantum state, was
proposed 60 years ago (\cite{Husimi}, see also \cite{Lee,Glaub,Sudar}). During the last decade
another very interesting representation has been actively developed: the quantum tomography,
operating with the ensemble of {\it scaled and rotated reference frames}, instead of the
phase space \cite{Man'ko,ManMan'ko,DodMan'ko,Man'koMan'ko,KV,SM97}. In the framework of this
formalism the state-describing function (called marginal distribution or quantum tomogram) is
real and nonnegative. The advantage is that the quantum tomogram is a {\it probability
distribution} shown to completely describe the quantum state \cite{SM95,GMD}.
It is one of the reasons why the quantum tomography has become so popular.

In this paper we propose a new method for computer simulation of nonstationary quantum processes
based on the tomography representation of quantum mechanics and illustrate it considering the
problem of nonstationary tunneling of a wave packet. Many simulation approaches in nonstationary
quantum mechanics are based on the numerical solution of the time-dependent Schr\"odinger
equation. There are also methods using the ensembles of classical trajectories to simulate
quantum evolution. For example, the method of "Wigner trajectories", based on the Wigner
representation \cite{Wigner}, is well known (see, {\it e.g.}, review \cite{Lee} for details) and
was recently successfully applied to investigate the tunneling of a wave packet \cite{DM,LF}.
Using such trajectories, one does not have to store the large arrays representing, say, the wave
function (contrary to grid methods), therefore there is a gain in computer memory.

The quantum tomogram $w$ depends on the variables $\{X,\mu,\nu\}$, where $X = \mu q + \nu p$,
and $q,p$ are the coordinates and momenta of the system, respectively, $\mu, \nu$ are the
parameters of scaling and rotation of reference frame in the phase space. The quantum tomogram
is nonnegative and normalized in $X$ direction, therefore it can be interpreted as a distribution
function of the value $X$. In our method the ensemble of trajectories in space $\{X,\mu,\nu\}$
is introduced to describe the quantum evolution. The trajectories are governed by the dynamical
equations obtained from the evolution equation for the quantum tomogram.

We demonstrate the method considering the nonstationary tunneling of a wave packet through the
potential barrier. For this problem we calculated tunneling times, which are of interest nowadays,
both for basic science ({\it e.g.}, what is the time spent by an atom to tunnel from the trap?)
and for applications (electronic tunneling time is connected with the operation rate of some
nanostructure-based devices). We also analyzed the evolution of the wave packet in coordinate and
momentum spaces in details. Another demonstration was designed to show that our method is not
restricted to one-particle simulations. Namely, we investigated the tunneling of a wave packet of
composite quasiparticle, an exciton (coupled electron and hole in semiconductor), in a
one-dimensional nanostructure (quantum wire). In this case we had two degrees of freedom. For
this problem, in addition to the probability density evolution, we determined the probability
of ionization due to electron and hole scattering on the barrier in different directions.

In Sec.~\ref{Method} we present description of the method, in Sec.~\ref{Tunneling} describe the
model problem and main results for a wave packet tunneling. Exciton tunneling is considered in
Sec.~\ref{Exciton} and the work is summarized in Sec.~\ref{Conclusion}.

\section{The method of simulation}\label{Method}
The quantum tomogram $w(X,\mu,\nu)$ is connected with the density matrix $\rho(q,q')$ as \cite{SMJ,VIM}:
\begin{eqnarray}
&& \rho(q,q') = \int w(X,\mu,q-q')e^{i(X - \mu (q+q')/2)}\frac{d\mu dX}{2\pi},\\ \label{WtoDM}
&& w(X,\mu,\nu) = \nonumber\\
&& \int e^{-i(k(X-\mu q-\nu p)+pu)}\rho(q+\frac u 2,q-\frac u 2)\frac{dp dk dq du}{2\pi^2}\label{DMtoW}
\end{eqnarray}

Consider the case of the particle with mass $m$ in one-dimensional space. If the Hamiltonian of the system is
\begin{equation}
 H = \frac{p^2}{2m}+V(q),
\end{equation}
then the integral transformation (\ref{DMtoW}) applied to the time-dependent
evolution equation for the density matrix gives \cite{Man'ko}
\begin{eqnarray}
 \dot w - \frac \mu m \frac{\partial w}{\partial \nu} - 2\frac{\partial V(\tilde q)}{\partial
 q}\left(\frac \nu 2 \frac{\partial}{\partial X}\right)w + \nonumber\\
 2\sum_{n=1}^\infty \frac{(-1)^{n+1}}{(2n+1)!}\frac{\partial^{2n+1}V(\tilde q)}{\partial
 q^{2n+1}}\left(\frac \nu 2 \frac{\partial}{\partial
 X}\right)^{2n+1}w = 0,
 \label{EvT}
\end{eqnarray}
where we use $\hbar = 1$ and $\tilde q$ is given by
\begin{equation}
 \tilde q = - \left(\frac{\partial}{\partial X}\right)^{-1} \frac{\partial}
 {\partial \mu}
 \label{OprQ}
\end{equation}

Eq.(\ref{EvT}) can be rewritten as
\begin{eqnarray}
 \frac{\partial w}{\partial t} + \frac{\partial w}{\partial X}
 G_X(X,\mu,\nu) + \frac{\partial w}{\partial \mu}G_{\mu}(X,\mu,\nu) +\nonumber\\
 \frac{\partial w}{\partial \nu}G_{\nu}(X,\mu,\nu) = 0,
 \label{ContEqXMuNu}
\end{eqnarray}
where functions $G$ depend on quantum tomogram, its derivatives and antiderivatives (the latter
corresponding to terms with $(\partial/\partial X)^{-1}$ in Eq.(\ref{EvT})). Generalization for
the case of more variables is straightforward because the form of the equations does not change.
Functions $G$ for the problem under investigation are given in Sec.~\ref{Tunneling}. The
evolution equation rewritten as (\ref{ContEqXMuNu}) has the form of continuity equation for the
quantum tomogram
\begin{equation}
 \frac{dw}{dt} = \frac{\partial w}{\partial t} + \frac{\partial w}{\partial X}
 \dot X + \frac{\partial w}{\partial \mu}\dot \mu + \frac{\partial w}{\partial \nu}\dot\nu = 0
 \label{ContEq}
\end{equation}
This equation is analogous to the continuity equation for classical distribution function
and Liouville equation. As known, the characteristics of Liouville equation are the classical
trajectories in phase space and they obey Hamilton equations of motion. The
quantum tomogram is nonnegative and we use it as a distribution function for trajectories
in the space $\{X, \mu, \nu\}$, obeying the equations analogous to Hamilton equations for
the classical trajectories. From the comparison of Eq.(\ref{ContEqXMuNu}) with Eq.(\ref{ContEq})
it is obvious that the trajectories are governed by the equations
\begin{equation}
 \dot X = G_X(X,\mu,\nu),
 \dot \mu = G_{\mu}(X,\mu,\nu),
 \dot \nu = G_{\nu}(X,\mu,\nu)
 \label{EqMotionG}
\end{equation}

The trajectories are used to avoid the direct calculations of the distribution function (contrary
to grid methods where the wave function is calculated at every point to solve numerically
Schr\"odinger equation). Hence it is necessary to use some approximation for the quantum tomogram
and we use local exponential approximation (as in Ref.~\cite{DM} for the Wigner function):
\begin{equation}
 w(X,\mu,\nu) = w_0e^{-[(y-y_a(t))A_a(t)
 (y-y_a(t))+b_a(t)(y-y_a(t))]},
 \label{TLocA}
\end{equation}
where $y = \{X, \mu, \nu\}$, and $y_a$ is the point under consideration. Parameters of this
approximation are matrix $A_a$ and vector $b_a$, and some combinations of these parameters enter
the evolution equation (\ref{EvT}), instead of the derivatives and antiderivatives of the quantum
tomogram. Calculation of average $X, \mu, \nu$ and their average products allows to obtain $A_a$
and $b_a$. After that, functions $G$ are known and dynamical equations (\ref{EqMotionG}) can be
solved numerically.

We would like to emphasize that we use the local approximation (\ref{TLocA}) only for the
calculation of r.h.s. in the equations of motion (\ref{EqMotionG}). The use of ensemble of trajectories to
represent the quantum tomogram means the approximation of quantum tomogram as a set of
delta-functions, each delta-function corresponds to one trajectory. If the number of trajectories
approaches infinity, the quantum tomogram can be approximated by the set of delta-functions with
arbitrary precision. This is analogous to what is conventionally done in classical statistical
mechanics for the distribution function in phase space. But unlike the classical statistical
mechanics now the trajectories are not independent: the approximation (\ref{TLocA}) is used to
take the non-local character of quantum-mechanical evolution into account.

The validity of this approximation holds if the quantum tomogram is smooth and the trajectories
are close to each other. For example, this approximation can fail if one tries to consider a
plain wave with wave vector $k$: in this case $w(X,\mu=0,\nu=1) = \delta(X-k)$. Approximation
(\ref{TLocA}) also works not well for unbounded motion, because the trajectories scatter with
time. If there are few trajectories in the region around the given point, then the approximation
(\ref{TLocA}) will not reconstruct the quantum tomogram well, due to lack of statistics.

We consider the tunneling of wave packets through the barrier, and comparing our results with
exact quantum computation we see that approximation (\ref{TLocA}) holds for this problem (see
Secs.~\ref{Tunneling},\ref{Exciton}). For example, considering the tunneling through the
potential barrier from the well, we deal with both the region of bounded motion (in the well) and
unbounded motion (beyond the barrier), still the approximation works not bad (Sec.~\ref{Tunneling}).
For higher initial energy the penetration through the barrier is higher, and the evolution of most
trajectories corresponds to unbounded motion, then the validity of Eq.(\ref{TLocA}) becomes poorer
(see the end of Sec.~\ref{Tunneling}), in agreement with the discussion above. But in general local
approximation (\ref{TLocA}) works satisfactory even for quite long time intervals (see Fig.~\ref{Fig1}).

To obtain any information about the system, we have to calculate some average values. Consider an
arbitrary operator $A(\opr q,\opr p)$. Average value $\langle A\rangle$ of corresponding physical
quantity is calculated in the tomographic representation of quantum mechanics as \cite{OM}
\begin{equation}
\langle A\rangle = \int A(\mu,\nu)e^{iX}w(X,\mu,\nu)dXd\mu d\nu,
\label{Aver}
\end{equation}
where $A(\mu,\nu)$ is the Fourier component of the Weyl symbol $A^W(q,p)$ of operator
$A(\opr q,\opr p)$ (see, {\it e.g.}, \cite{Lee}):
\begin{equation}
A(\mu,\nu) = \int A^W(q,p)exp(-i(\mu q + \nu p))\frac{dqdp}{4\pi^2}\label{FofW}
\end{equation}

For the calculation of average values we use the following approximation of quantum tomogram:
\begin{equation}
w(X,\mu,\nu,t) = \sum_{j=1}^J\delta(X-X_j(t))\delta(\mu-\mu_j(t))
\delta(\nu-\nu_j(t)),\label{TomAp}
\end{equation}
where the summation is made over all $J$ trajectories; $X_j(t),\mu_j(t),\nu_j(t)$ are the
coordinates of the $j$-th trajectory in $\{X,\mu,\nu\}$ space at time $t$. Such approximation
corresponds to use of the ensemble of trajectories. In the regions, where $w(X,\mu,\nu)$ is
small, trajectories are rare, and where it is large, trajectories are accumulated. The more
trajectories are used, the better the approximation (\ref{TomAp}) works. If during the simulation
the wave function has the form of a compact wave packet, even consisting of several distinct parts,
approximation (\ref{TomAp}) holds, because in this case one has the compact sets of trajectories
providing good statistics. This is the case for problems considered, and therefore the use of
this approximation does not change results essentially, in comparison with the exact quantum
computation (Secs.~\ref{Tunneling},\ref{Exciton}).

For the operators $A(\opr q)$, depending on $\opr q$ only, expression for $\langle A\rangle$
takes the form:
\begin{equation}
\avr A = \int A(X)w(X,\mu=1,\nu=0)dX,\label{AverAq}
\end{equation}
where $A(X)$ is the function corresponding to the operator $A(\opr q)$ in coordinate
representation, $A(X) = A(q = X)$. The method of calculation of an average $\avr{A(\opr q)}$ at
arbitrary time $t$, with the approximation (\ref{TomAp}), is quite simple. One just takes into
account the trajectories with any $X$ and with $\mu(t), \nu(t)$ from the small region near
$\mu = 1, \nu = 0$ only, and performs a summation of $A(X)$ over all such trajectories.

The developed method is similar to well-known method of Wigner trajectories (see \cite{Lee} for
review), where the ensemble of trajectories is introduced in the phase space, with the Wigner
function used as a quasi-distribution function. The quantum tomogram is defined in the space
$\{X,\mu,\nu\}$, that is not as simple for understanding as the phase space used in the Wigner
approach. On the other hand, quantum tomogram is true distribution function, while the Wigner
function can be both positive and negative. In spite of these differences the two approaches are
quite close in general, and there are some difficulties common for both methods. The discussion
(see above) of the approximation for the tomogram (such as (\ref{TLocA})) concerns, in principle,
the method of Wigner trajectories either. Another important example is the discontinuity of the
Wigner trajectories discussed in Ref.~\cite{SBMJChemPhys1993}, that is due to the fact that the
trajectories are not independent as in classical statistical mechanics, their evolution depends on
their local distribution. The same is applicable for the quantum tomography approach, where the
trajectories are also not independent for the same reason. In Ref.~\cite{SBMJChemPhys1993} a possible
alternative to Wigner function was proposed. Namely, in that work the authors proposed to use the
Weyl transforms of some operators (the Wigner function, up to the constant, is the Weyl transform
of the density operator) instead of the Wigner function to generate the ensemble of trajectories.
The same can be introduced for the quantum tomography. Applying the transform (\ref{DMtoW}) to the
matrix elements of an operator $A$ we obtain the symbol $w_A(X,\mu,\nu)$. This function is not
nonnegative in general, but in analogy with the Wigner trajectories one can use $w_A(X,\mu,\nu)$
to develop some new ensemble of trajectories. Probably, as with the Weyl transforms trajectories
\cite{SBMJChemPhys1993}, it will be more convenient to use the trajectories corresponding to
$w_A$ in certain cases. Of course, this problem needs further investigation.

\section{Simulation of tunneling of a wave packet}\label{Tunneling}
\subsection{The model and calculated average values}
We choose the external potential to coincide with the potential used in \cite{DM}, for comparison
of the results of simulation in quantum tomography approach with those obtained by other methods.
We consider behavior of the wave packet in one-dimensional space in external potential
\begin{equation}
 V(q) = \frac{m\omega_0^2q^2}{2} - \frac{bq^3}{3}
 \label{Vext}
\end{equation}
As the potential has only the second and third powers of coordinate, all its derivatives of
order more than the third vanish. Evolution equation in this case has the form ($\hbar = 1$):
\begin{eqnarray}
 \frac{\partial w}{\partial t} - \frac \mu m \frac{\partial w}{\partial \nu} +
 2\left[-\frac{\partial V(\tilde q)}{\partial q}\left(\frac \nu 2 \frac{\partial}{\partial X}\right)+\right.\nonumber\\
 \left.\frac 1 6 \frac{\partial^3V(\tilde q)}{\partial q^3}
 \left(\frac \nu 2 \frac{\partial}{\partial X}\right)^3\right]w = 0
 \label{EvTX23}
\end{eqnarray}
For the potential given by Eq.(\ref{Vext}) evolution equation reads as:
\begin{eqnarray}
 \frac{\partial w}{\partial t} - \frac \mu m \frac{\partial w}{\partial \nu}
 + m\omega_0^2\nu\frac{\partial w}{\partial \mu} -
 \frac{b \nu^3}{12}\frac{\partial^3 w}{\partial X^3} +\nonumber\\
 b\nu\left(\frac{\partial}{\partial X}\right)^{-1}\frac{\partial^2
 w}{\partial \mu^2} = 0,
 \label{EvTX23Cont}
\end{eqnarray}
and dynamical equations have the form
\begin{eqnarray}
 \frac{\partial X}{\partial t} &=& \frac{b\nu^3}{12} \frac
 1 w \frac{\partial^2 w}{\partial X^2}\nonumber\\
 \frac{\partial \mu}{\partial t} &=& m\omega_0^2\nu - \frac{b\nu}{w}
 \left(\frac{\partial}{\partial X}\right)^{-1}\frac{\partial w}{\partial \mu}\nonumber\\
 \frac{\partial \nu}{\partial t} &=& - \frac \mu m
 \label{EqMotionM}
\end{eqnarray}

We use atomic units throughout, $\hbar = m_e = |e| = 1$, where $m_e$ and $e$ are the mass and
charge of a free electron. The particle with mass $m = 2000$ is regarded. Parameters of
the potential are $\omega_0 = 0.01$ and $b = 0.2981$. This potential has the minimum at
$q = 0$ $(V(0) = 0)$ and maximum at $q = 0.6709$ $(V(0.6709) = 0.015)$, therefore here we
consider the motion of a particle in the potential well with infinite left wall and the
barrier of height $0.015$ at $q = 0.6709$. This model problem roughly describes
nonstationary tunneling of an atom from the trap.

Initially the particle represented by the wave packet is situated to the left from $q = 0$, its
mean momentum is zero. The particle can oscillate in the potential well and can tunnel or pass
above the barrier. The probabilities of these processes depend on the initial energy of the wave
packet. We consider the problem, where all parameters, except the initial mean coordinate $q_0$
of the wave packet, are fixed (initial mean momentum equals zero, dispersions of the wave packet
in coordinate and momentum spaces are $\approx 0.3$ and $\approx 1.6$, respectively).

We solve the equations (\ref{EqMotionM}) numerically. As in \cite{DM} we consider three values
of $q_0$: $-0.2, -0.3$ and $-0.4$. The most interesting quantities characterizing tunneling
are reaction probability and tunneling time. Reaction probability is defined as
\begin{equation}
\int\limits_{q_a}^{\infty}|\psi(x,t)|^2dx, \label{ReactProb}
\end{equation}
where $q_a = 0.6709$ (the point where potential has the maximum), the maximum value of reaction
probability is unity. Reaction probability shows what part of the wave packet is currently beyond
the barrier.

Tunneling time of the wave packet is also an important feature of tunneling. There are a lot of
methods to determine tunneling time \cite{BL1982,LM,SokConnor,Baz',Ryb,But,HS,BL,AB,Wlod,BEMuga,
LLBR,DelgMuga,MLPhysRep2000}. We use the approach where tunneling time is calculated as the
difference of {\it presence times} (see \cite{MLPhysRep2000} for review) at point $x_a$ and $x_b$,
located on the opposite sides of the barrier:
\begin{equation}
t_T(x_a,x_b) = \langle t(x_b)\rangle-\langle t(x_a)\rangle\label{TunTime}
\end{equation}
The presence time at arbitrary point $x_0$ is
\begin{equation}
\langle t(x_0)\rangle = \frac{\int\limits_0^\infty t |\psi(x_0,t)|^2dt}
{\int\limits_0^\infty |\psi(x_0,t)|^2dt}\label{PresTime}
\end{equation}

\subsection{Reaction probability}
\begin{figure}
 \includegraphics[height=7cm]{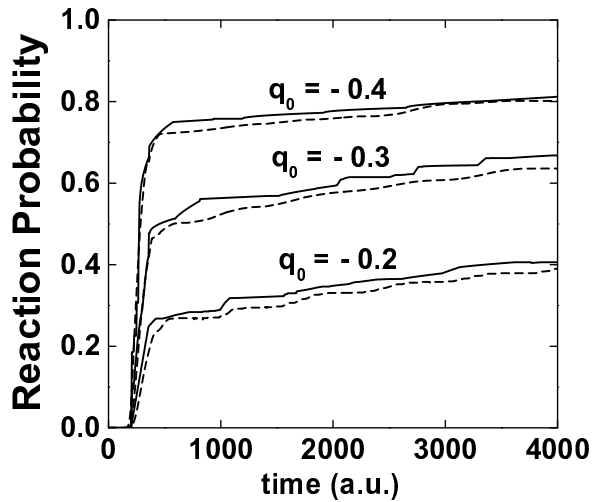}
 \caption{\label{Fig1}The dimensionless reaction probabilities (\ref{ReactProb}) for three
 values of initial mean coordinate of the wave packet: $q_0 = -0.2$, $-0.3$, and $-0.4$ a.u.
 Solid lines are for the simulation in quantum tomography approach, dashed lines are
 for the exact numerical solution.}
\end{figure}

Here we present the results obtained by means of our method and compare them with the exact
numerical solution of Schr\"odinger equation. In Fig.~\ref{Fig1} we present the dependence of
reaction probability (\ref{ReactProb}) on time for three values of initial mean coordinate of the
wave packet $q_0 = -0.2, -0.3$ and $-0.4$, corresponding mean energies of the wave packet are
$\approx 0.75V_0$, $\approx 1.25V_0$ and $\approx 2.0V_0$, respectively. Solid lines represent
the results of simulation in the quantum tomography approach (QT) and dashed lines correspond to
the numerical solution of Schr\"odinger equation (exact quantum computation). Due to the increase
of initial mean energy with the increase of $|q_0|$, the portion of high energy components in the
wave packet grows. This leads to the higher portion of components, which pass through the barrier,
either because their energy is greater than the height of the barrier, or due to tunneling.
Therefore, with the growth of $|q_0|$, reaction probability becomes larger and one can see that
the curves corresponding to different $q_0$ are above each other in Fig.~\ref{Fig1}. The time
evolution of reaction probability is qualitatively the same for every $q_0$. The components,
which have passed through the barrier, can not return, because for $q > 0.6709$ potential
diminishes with the growth of coordinate, and so reaction probability can not decrease with
time. At first it grows rapidly due to transmission of components with the energy higher than
the height of the barrier (comparison with the classical solution of corresponding problem,
for which only transmission above the barrier is possible, convinces us in it). Then
reaction probability continues to grow slowly, because of the tunneling. All these features
present both for QT simulation and for exact quantum computation.

In comparison with the exact computation, reaction probability for the QT simulation is slightly
higher. Note also some difference in the character of increase of the reaction probability for QT
simulation and exact solution: in the former case the curves are not so smooth. These
differences are due to the finite number of trajectories used in QT simulation: for smaller
number of trajectories (not shown) reaction probability curves resemble staircase more evidently
(this is connected with the overestimation of the role of wave packet oscillations in the well
for the finite number of trajectories), and quantitative deviation from exact result is stronger.
But in general, for quite large number of trajectories, as for the case shown in Fig.~\ref{Fig1},
QT simulation results on reaction probability are quite close to those obtained through the exact
quantum computation (compare also with the method of "Wigner trajectories" in the work by
Donoso and Martens \cite{DM}).

\subsection{Evolution of the wave packet and tunneling times}
Besides the reaction probability we also obtained a number of new qualitative and quantitative
results, which described in details the behavior of the wave packet during tunneling. We also
calculated tunneling times using the concept of presence time (see below).

\begin{figure}
 \includegraphics[height=7cm]{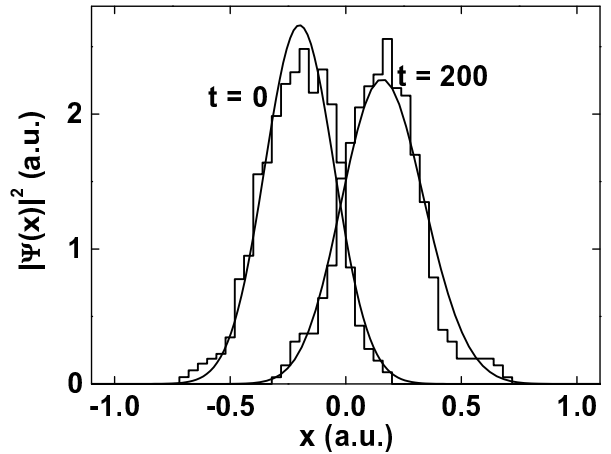}
 \caption{\label{Fig2} Probability density in coordinate space for QT simulation (histograms)
 and exact solution (smooth lines), at times $t = 0$ a.u. (left) and $t = 200$ a.u. (right). The
 barrier is at the point $0.6709$ a.u., $q_0 = -0.2$ a.u.}
\end{figure}

\begin{figure}
 \includegraphics[height=7cm]{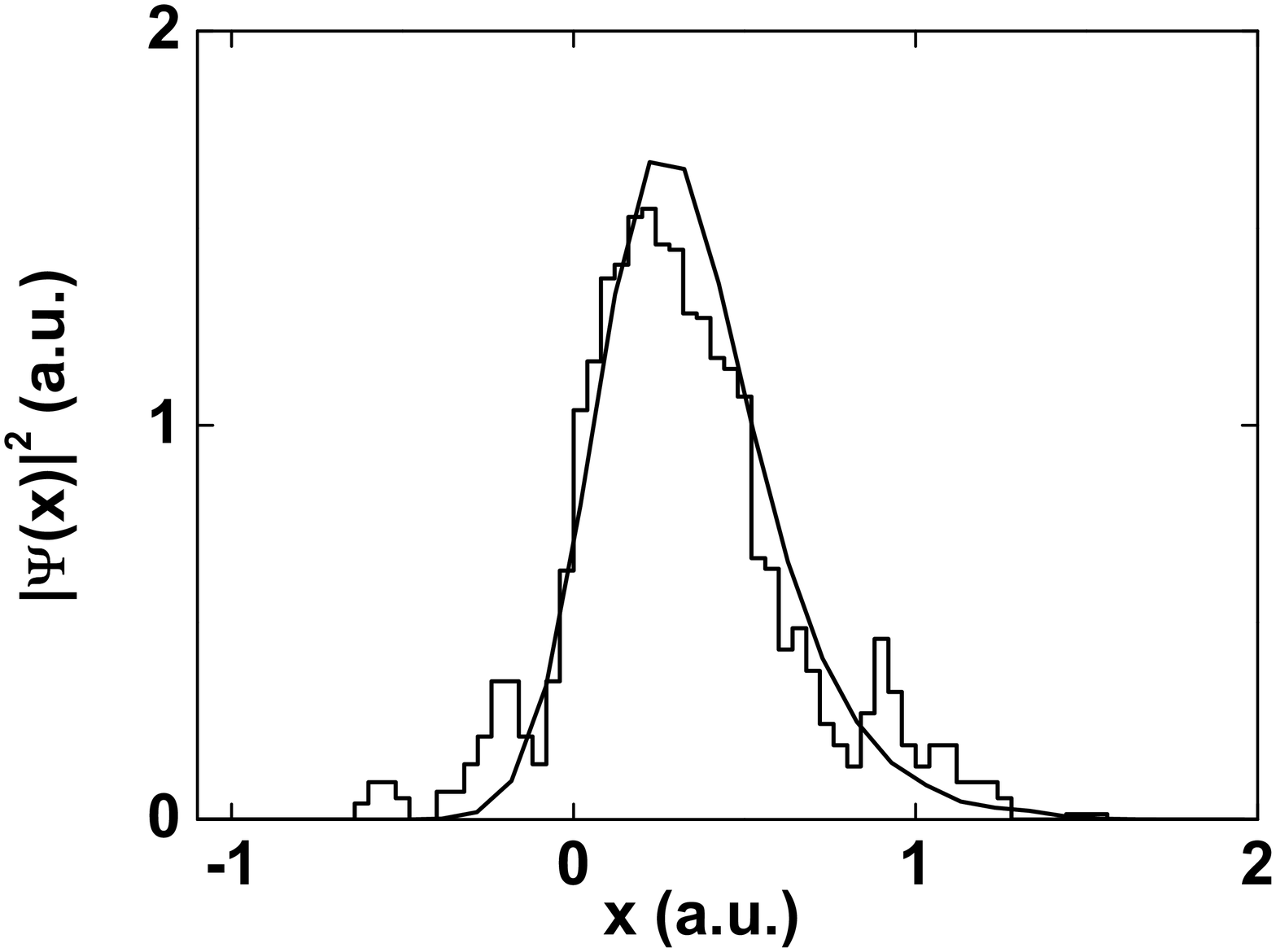}
 \caption{\label{Fig3} Probability density in coordinate space for QT simulation (histogram)
 and exact solution (smooth line), at time $t = 300$ a.u. The barrier is at the point $0.6709$
 a.u., $q_0 = -0.2$ a.u.}
\end{figure}

\begin{figure}
 \includegraphics[height=7cm]{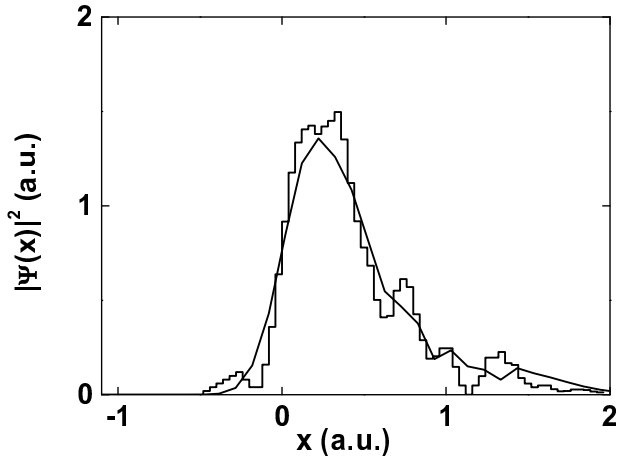}
 \caption{\label{Fig4} Probability density in coordinate space for QT simulation (histogram)
 and exact solution (smooth line), at time $t = 400$ a.u. The barrier is at the point $0.6709$
 a.u., $q_0 = -0.2$ a.u.}
\end{figure}

\begin{figure}
 \includegraphics[height=7cm]{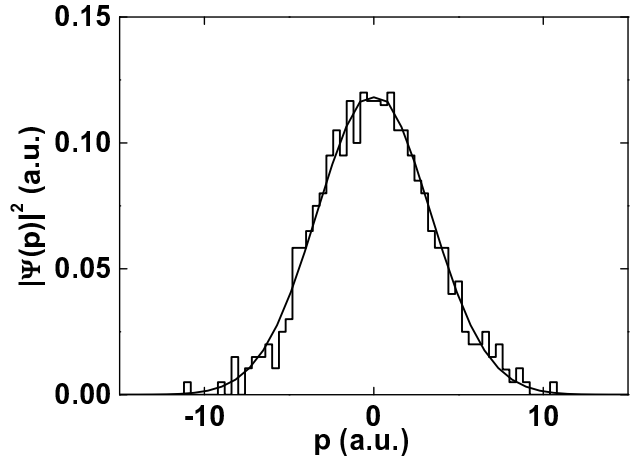}
 \caption{\label{Fig5} Initial probability density in momentum space for QT simulation
 (histogram) and exact solution (smooth line). $t = 0$ a.u.}
\end{figure}

\begin{figure}
 \includegraphics[height=7cm]{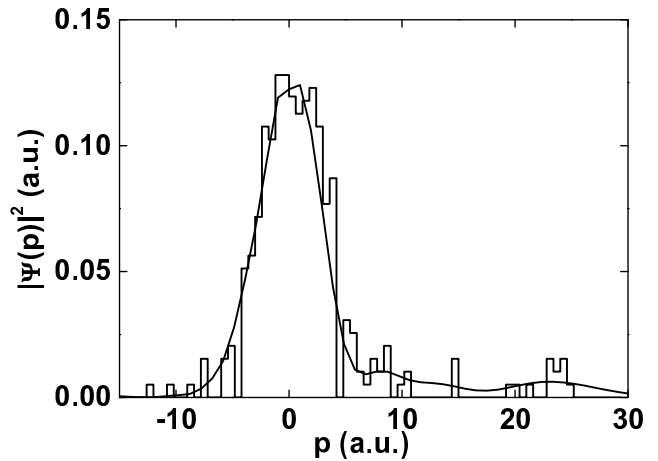}
 \caption{\label{Fig6} Probability density in momentum space for QT simulation (histogram) and
 exact solution (smooth line) at $t = 400$ a.u.}
\end{figure}

The following discussion concerns tunneling of the wave packet with initial mean coordinate
$q_0 = -0.2$. Note that the maximum of the potential is at the point $x = 0.6709$. We present
the normalized probability density $|\psi(x)|^2$ in coordinate space
(Figs.~\ref{Fig2}~-~\ref{Fig4}) and $|\psi(p)|^2$ in momentum space
(Figs.~\ref{Fig5}~,~\ref{Fig6}) for several successive time moments. In these figures smooth
lines show the shape of the wave packet obtained by means of exact quantum computation.
Histograms represent the result of {\it single QT run}. One can consider many runs with the same
number of trajectories and average the probability density over all these runs to obtain smoother
picture. But here we would like to show what QT simulation can give for one run, in comparison
with the exact quantum computation. Therefore the histograms (QT) fit the smooth solid lines in
Figs.~\ref{Fig2}~-~\ref{Fig4} (exact solution) not ideally, still the resemblance is obvious.

First, consider the probability density $|\psi(x)|^2$ in
coordinate space (Figs.~\ref{Fig2}~-~\ref{Fig4}). One can see
(Fig.~\ref{Fig2}) that initially the wave packet has Gaussian
form. It begins to move as a whole towards the potential minimum
at $x = 0$ (initial mean momentum is zero, but the potential
inclines in that direction), passes that point, is accelerated,
and collides with the barrier. During the motion the wave packet
broadens (due to dispersion in momentum space, compare right and
left plots in Fig.~\ref{Fig2}) but the interaction with the
barrier changes its form more substantially ($t = 300$ and $t =
400$, Figs.~\ref{Fig3},~\ref{Fig4}). The wave packet shrinks a
little, some components pass through the barrier and transmitted
part can be seen beyond the barrier ($x = 0.6709$). As the
transmitted part can not return and accelerates (potential
diminishes with the distance for $x > 0.6709$), the enriching of
the wave packet by high-energy components must present (see
below).

All features described in the previous paragraph are common for
both the exact solution and QT simulation. The histograms in
Figs.~\ref{Fig2}~-~\ref{Fig4} fit the smooth lines representing
exact solution better for earlier times, but even after the
interaction with the barrier (Fig.~\ref{Fig4}), resemblance is
quite close. This shows that approximations (\ref{TLocA}) and
(\ref{TomAp}) work not bad for the problem under consideration.

Proceed now to the evolution of the wave packet in momentum space. To
confirm our analysis on the acceleration of the transmitted part
of the wave packet, we present the probability density
$|\psi(p)|^2$ in momentum space in Figs.~\ref{Fig5},~\ref{Fig6},
at times $t = 0$ and $t = 400$, respectively. As the wave
function is initially the Gaussian wave packet, the initial
distributions both in coordinate and momentum space are Gaussian
(compare Fig.~\ref{Fig2} and Fig.~\ref{Fig5}). But after the wave
packet has interacted with the barrier, the distribution in momentum
space changes substantially (Fig.~\ref{Fig6}). The higher the
energy of incident particle, the greater the tunneling
probability. Therefore the barrier transmits mainly the wave
packet components with relatively high energy, serving as an
energy selector. Components with high momentum arise, as the
transmitted part of the wave packet is accelerated in the region
of lowering potential beyond the barrier, and we
observe the enriching of the wave packet by high energy
components. The resemblance between the histograms (QT
simulation) and smooth solid lines (exact solution) is somewhat
poorer for momentum distribution at large times ($t = 400$,
Fig.~\ref{Fig6}) than for coordinate distribution
(Fig.~\ref{Fig4}). This is due to the fact that one deals with
finite number of trajectories and has to sample quite large
interval in momentum space, because the transmitted part is
permanently accelerated. Therefore, with time, momentum
distribution spreads and there are not many
trajectories with $\mu, \nu$ close enough to $\mu = 0, \nu = 1$
for given momentum $p$ (see Sec.~\ref{Tunneling}). As for the
considered value of initial mean coordinate $q_0 = -0.2$ the initial
energy is not very large ($\approx 0.75V_0$, where $V_0$ is the
height of the barrier), the wave packet mainly stays in the well
(for the time considered $t = 400$ only $\approx 20\%$ of the
wave packet is transmitted, see Fig.~\ref{Fig1}) and the
distribution in coordinate space is more compact.

\begin{figure}
 \includegraphics[height=7cm]{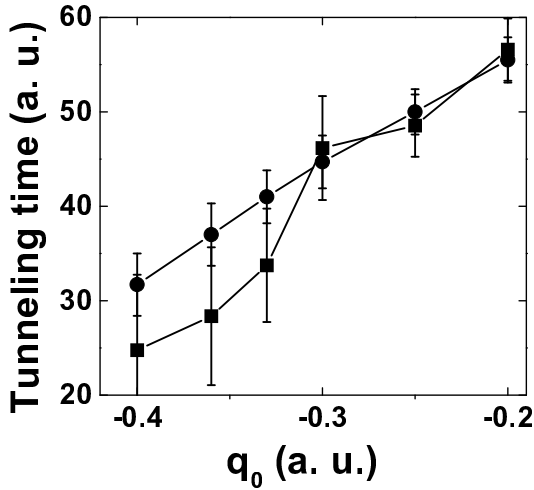}
 \caption{\label{Fig7} Tunneling times with errors for several values of initial mean coordinate
 of the wave packet $q_0$. Results of the QT simulation (squares) are compared with exact quantum
 computation (circles).}
\end{figure}

In Fig.~\ref{Fig7} we present the dependence of tunneling time on
initial mean position of the wave packet. Tunneling time is
determined as the difference of presence times (\ref{PresTime})
for points $x_a = 0.5\cdot0.6709$ and $x_b = 2.0\cdot0.6709$ (at
$x = 0.6709$ potential has the maximum). Tunneling is usually
stronger for the higher energy. Therefore, the increase of
$|q_0|$ (and corresponding increase of initial mean energy) leads
to the growth of the average speed both of the transmitted part
and of the wave packet as a whole. The transmitted part passes
the region of the barrier (the space between the points $x_a$ and
$x_b$) faster, and so one expects that the increase of $|q_0|$
causes the decrease of tunneling time. Indeed, the value of
tunneling time drops with increase of $|q_0|$. Results of QT
simulation (squares in Fig.~\ref{Fig7}) deviate from those of
exact computation (circles) within the range of errors. The
deviation is maximal for large $|q_0|$. Probably, this is because
for large $|q_0|$ the wave packet leaves the well almost entirely
(see Fig.~\ref{Fig1}), and the evolution of most trajectories
corresponds to the unbounded accelerated motion. In such
situation trajectories scatter and approximation
(\ref{TomAp}) does not represent the quantum tomogram as exactly
as for smaller $|q_0|$.

\section{Simulation of the exciton tunneling}\label{Exciton}
The method described in Sec.~\ref{Method} can be used to simulate the evolution of systems with
more than one degree of freedom. In this section we demonstrate such a possibility, considering
nonstationary tunneling of the composite particle, exciton, through the potential barrier in
one-dimensional (1D) semiconductor structure (quantum wire). Exciton is a bound state of electron
and hole in semiconductor, therefore we deal with two degrees of freedom in contrast to
Sec.~\ref{Tunneling}.

Possible experimental realization is as follows. Consider semiconductor quasi-one-dimensional
nanostructure where the motion is allowed only in one direction (quantum wire). Transverse motion is
restricted due to strong confining barriers. Potential barrier in the direction of allowed motion
can be located at some point of the quantum wire either using the semiconductor heterojunction or
by the gate. Using femtosecond laser pulses, we can form an excitonic wave packet, either by the
quasiresonance pumping, or exciting an electron from valence band with formation of a hole and
subsequent binding of two particles into exciton. Then the excitonic wave packet can move to the
barrier and with the help of some detectors one can investigate scattering of the exciton.

Keeping this in mind let us construct the model for the simulation. We use the constants
corresponding to GaAs for reference (dielectric constant $\varepsilon=12.5$, effective masses of
electron and hole are $m_e=0.07m_e^{(0)}$ and $m_h=0.15m_e^{(0)}$, respectively, where $m_e^{(0)}$
is the electron mass in vacuum). 3D exciton in bulk GaAs is characterized by effective Bohr radius
$a^*\approx 10 nm$ and binding energy $E_C^*\approx 4 meV$. We use unit of length $a^*$, unit of
mass $m_e$, and $\hbar=1$. Corresponding units of energy and time are
$E_0=\hbar^2/(m_ea^2)\approx10meV$ and $t_0 = m_ea/\hbar\approx100fs$.

The energy spectrum and wave functions of relative electron and hole motion in 3D exciton are
analogous to those of hydrogen atom. But this is not the case for 1D exciton. First, electron-hole
effective interaction potential in quasi-1D structure is not Coulomb. Indeed, if the exciton size in
the direction of allowed motion is much greater than the width of the quantum wire (in the transverse
direction), then the adiabatic approximation is applicable and 3D interaction potential must be
averaged over the transverse degrees of freedom. Resulting 1D effective potential substantially
differs from Coulomb (see \cite{LozFilA} for the discussion of similar model). Second, corresponding
energy spectrum and wave functions of electron and hole relative motion also change in comparison
with the hydrogen-like states. We choose the wave function of the exciton ground state in Gaussian form.

Excitonic wave packet can be represented as a Gaussian wave packet in center-of-mass coordinates:
\begin{equation}
\Psi(x_e,x_h,t=0)=\frac{e^{-r^2/(2\sigma)}}{(\pi\sigma)^{1/4}}\frac{e^{-(R-x_0)^2/(2S)+iRp_0}}{(\pi
S)^{1/4}},\label{InExcn}
\end{equation}
where $R=(m_ex_e+m_hx_h)/(m_e+m_h)$, $r=|x_e-x_h|$, $x_e$ and $x_h$ are electron and hole
coordinates, $x_0, p_0$ and $S$ are parameters; for them we used the following values:
$x_0=-10$, $p_0=3$, $S=2$ and $\sigma = 1$.

External potential is assumed to be zero everywhere except the region of barrier; we use the
barrier of thickness equal to $5nm$, or $0.5$ in accepted units. For simplicity we set the barriers
for electron and hole to be the same and use both external and interaction potentials in quadratic
form, cut at some distance. Then external potential is given by
\begin{equation}
V_{ext}(x)=
\begin{cases}
C-Dx^2,  & \text{if $|x| < \sqrt{\frac{C}{D}}$}, \\
0, & \text{if $|x| \ge \sqrt{\frac{C}{D}}$}\label{VExcnExt}
\end{cases}
\end{equation}
$C$ is the height of the barrier, its width is $\sqrt{C/D}=0.5$.

Interaction potential $V_{int}$ is also assumed to be quadratic:
\begin{equation}
V_{int}(r)=
\begin{cases}
Br^2-A,  & \text{if $r < \sqrt{\frac{A}{B}}$}, \\
0, & \text{if $r \ge \sqrt{\frac{A}{B}}$},\label{VExcnInt}
\end{cases}
\end{equation}
where $r=|x_e-x_h|$. Potential (\ref{VExcnInt}) can describe, {\it e.g.}, e-h interaction in
spatially indirect exciton, for example in coupled quantum wires with large interwire separation
\cite{LozWilAPA2000}. Initial wave function of relative motion, chosen to be Gaussian with
unity dispersion, is negligible within one percent accuracy at $r=3$. Thus we choose the radius of
electron-hole interaction to be $\sqrt{A/B}=3$.

We assume that we deal with a quasi-1D exciton with binding energy $E_C = 1/8$. In fact, for an
exciton in quantum wire, the wave function, binding energy, etc., are essentially influenced by the
properties of quantum wire. We also neglect the possibility of electron and hole recombination at
the time scales studied.

Here we consider just an example, therefore use relatively simple model. Still this model contains
the main features of exciton tunneling, such as the possibility of ionization, barrier and
interaction with realistic strength and size, the fact that composite particle is the bounded state
of two particles.

For stationary state the binding energy is $-E_C = \int\Psi^*_{int}(r)H_{int}(r)\Psi_{int}(r)dr$,
where $\Psi_{int}(r)$ is the wave function of relative motion. Then, from Eq.(\ref{VExcnInt}) and
condition $\sqrt{A/B}=3$ we have $A\approx 18E_C/17$.

We do not use the variables of relative and center-of-mass motion, for QT simulation it is easier to
deal with the initial conditions and evolution equation in coordinates of electron and hole $x_e$ and
$x_h$. Potentials (\ref{VExcnExt}) and (\ref{VExcnInt}) are quadratic, that makes tomographic
consideration of the problem easier (see Sec.~\ref{Tunneling}), discontinuity is neglected. On the
other hand, in coordinates $x_e$ and $x_h$ the evolution equations depend on trajectory distribution
(see Sec.~\ref{Method}), therefore the problem considered allows to employ all techniques, developed
for one degree of freedom, in this case of two degrees of freedom.

\begin{figure}
 \includegraphics[height=7cm]{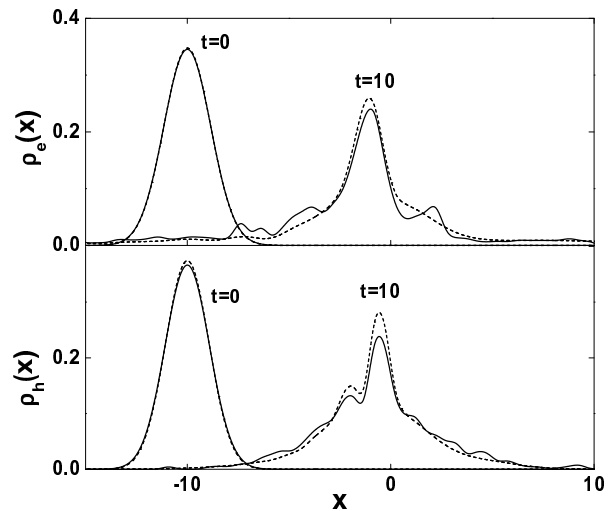}
 \caption{\label{Fig8} Probability density distributions in coordinate space for electron
 ($\rho_e(x)$) and hole ($\rho_h(x)$) at times $t=0$ and $t=10$. QT simulation (solid lines) is
 compared with exact numerical solution (dashed lines). All values are in units $\hbar=m_e^*=E_C
 =1$, $m_e^*$ is electron effective mass and $E_C$ is binding energy of the exciton. The height of
 the barrier $C = 1$, width $\sqrt{C/D} = 0.5$.}
\end{figure}

\begin{figure}
 \includegraphics[height=7cm]{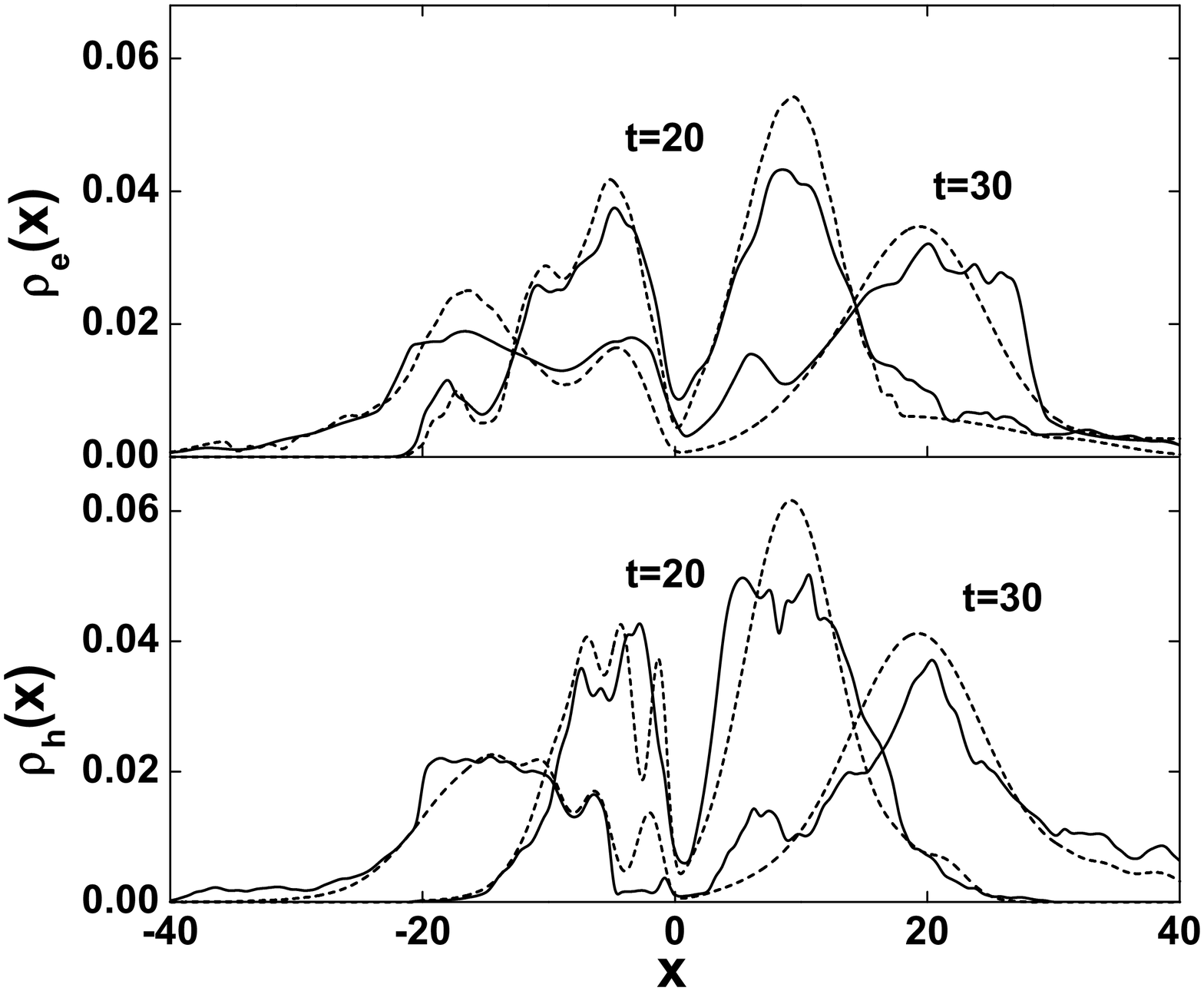}
 \caption{\label{Fig9} Probability density distributions in coordinate space for electron
 ($\rho_e(x)$) and hole ($\rho_h(x)$) at times $t=20$ and $t=30$. QT simulation (solid lines) is
 compared with exact numerical solution (dashed lines). The same units and barrier parameters
 as in Fig.~\ref{Fig8} are used.}
\end{figure}

\begin{figure}
 \includegraphics[height=7cm]{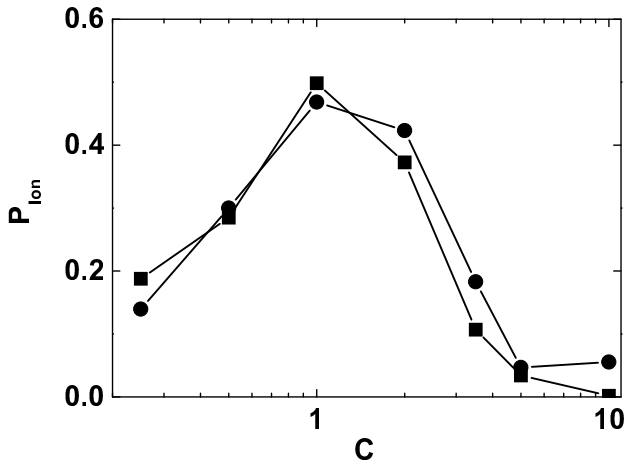}
 \caption{\label{Fig10} Probability of exciton ionization $P_{Ion}$ due to electron and hole scattering
 on the barrier in opposite directions {\it versus} barrier height $C$. Circles and
 squares represent QT simulation and exact solution, respectively. Considered is the barrier of
 thickness $0.5$. The units are the same as in Fig.~\ref{Fig8}.}
\end{figure}

Results of exciton tunneling simulation are presented in Figs.~\ref{Fig8}-\ref{Fig10}. All
parameters are fixed (see above), except the barrier height $C$ in Fig.~\ref{Fig10}. In
Figs.~\ref{Fig8} and \ref{Fig9} we depict the evolution of probability density for electron and
hole, barrier height $C = 1$. Solid lines in Figs.~\ref{Fig8}~-~\ref{Fig9} and circles in
Fig.~\ref{Fig10} correspond to the simulation in quantum tomography approach, while dashed lines and
squares represent the exact numerical computation. Unlike in Figs.~\ref{Fig2}~-\ref{Fig6}, here we
show the results of several combined QT simulation runs, therefore corresponding lines are
relatively smooth. Coincidence of QT and exact computation results, initially very good, becomes
poorer with time (Figs.~\ref{Fig8} and \ref{Fig9}), but QT simulation reproduces main properties of
exciton tunneling: wave packets broadening with time and due to interaction with the barrier,
shrinking near the barrier, dividing into two parts (reflected and transmitted). Note that QT
results are quite close to exact ones even for long times ($t=30$), despite the fact that the motion
is unbounded (see Sec.~\ref{Method}). Integral values (in Fig.~\ref{Fig10}) obtained in QT approach
also agree with the exact results. Larger discrepancies correspond to higher barriers, probably due
to larger inaccuracies, introduced by neglecting the potential discontinuity in the case of
stronger interaction with the external potential.

The electron and hole wave packets begin the motion from the point $x=-10$ and, shrinking near the
barrier, are partially reflected and transmitted. For the case presented in Figs.~\ref{Fig8}~-~
\ref{Fig9} about the half of wave packets is transmitted. Interesting is the question about the
ionization probability of exciton, induced by interaction with the barrier. If electron and hole are
scattered in different directions on the barrier, the distance between them can become quite large,
but, in principle, there is a possibility that exciton is not ionized after such scattering, because
one of the particles can be 'pulled' beyond the barrier, to the other particle, due to electron-hole
attraction. On the other hand, the electron-hole interaction is cut at the distance $\sqrt{A/B}$ in
our model. After the interaction with the barrier the wave packet divides into reflected and
transmitted parts moving in opposite directions. For the time large enough, these two parts are well
separated, the separation between them grows and the leakage through the barrier in both directions
is negligible. Denote the probability of ionization due to electron and hole scattering in different
directions as $P_{Ion}$. Then, the probability to find electron and hole in different directions in
respect to the barrier, with e-h distance being larger than $\sqrt{A/B}$, approaches $P_{Ion}$ in
the limit $t\rightarrow\infty$.

The probability of ionization due to electron and hole scattering in different directions on the
barrier $P_{Ion}$ is presented in Fig.~\ref{Fig10}, depending on the barrier height $C$. For very
high and very low barriers $P_{Ion}$ must approach zero, because in former case both particles are
reflected and in the latter they both are transmitted. This trend is seen in Fig.~\ref{Fig10}, and
$P_{Ion}$ depending on $C$ is maximal (other parameters are fixed, see above) at $C\approx 1$. Note
that these features are obvious for curves representing both QT simulation (circles) and exact
computation (squares), and in general two curves are quite close to each other.

\section{Conclusion}\label{Conclusion}
We have developed the new method of numerical simulation of quantum nonstationary processes
and applied it to the problem of tunneling of the wave packet through the potential barrier. The method is
based on tomographic representation of quantum mechanics. The quantum tomogram is used in a sense as the
distribution function for the ensemble of trajectories in space $X,\mu,\nu$, where $X = \mu q + \nu p$ is the
coordinate measured in rotated and scaled reference frame, $q,p$ are coordinate and momentum of the system,
respectively. The trajectories are governed by the equations, resembling the Hamilton
equations of motion, therefore, some analogue of molecular dynamics can be used. The
Gaussian approximation allows to avoid the direct calculation of quantum tomogram. Instead
of quantum tomogram, the parameters of the approximation are used in the equations of
motion. Those parameters can be obtained if one calculates the local moments of the
ensemble of the trajectories.

The problem of nonstationary tunneling of the wave packet was considered. Our method gave the
results in agreement with those obtained by the method of "Wigner trajectories" and by exact
quantum computation.

Of course, we made only the first step in the development of this simulation method, having
considered one-dimensional problem. But it is obvious that the generalization for the case of
several dimensions is straightforward. In this case the method possesses the additional advantage
of more rapid convergence due to the fact the quantum tomogram is nonnegative. This may help to
overcome the problem of sign for fermionic systems. In the next work we intend to
consider the many-body problem for fermionic and bosonic systems by means of the quantum
tomography method. The similar approach (in the framework of "Wigner trajectories") has been
already applied for the investigation of the tunneling of two identical particles \cite{LozFilA}.

\begin{acknowledgments}
Authors are grateful to INTAS, RFBR and Ministry of Science for
financial support. A. A. also acknowledge the financial help of
the Dynasty foundation and ICFPM. We thank V. I. Man'ko for
fruitful discussions.
\end{acknowledgments}
\appendix

\end{document}